\documentclass{article}
\usepackage{amssymb}
\usepackage{amsmath}
\usepackage{amsfonts}
\usepackage{rotating}
\usepackage{subfig}
\usepackage{graphicx}
\usepackage[all]{xy}
\usepackage{color}
\usepackage{verbatim}
\usepackage{natbib}
\usepackage{prodint}%
\usepackage{authblk}
\providecommand{\U}[1]{\protect\rule{.1in}{.1in}}

\definecolor{gray}{rgb}{0.66, 0.66, 0.66}
\ifx\pdfoutput\relax\let\pdfoutput=\undefined\fi
\newcount\msipdfoutput
\ifx\pdfoutput\undefined\else
\ifcase\pdfoutput\else
\ifx\paperwidth\undefined\else
\ifdim\paperheight=0pt\relax\else\pdfpageheight\paperheight\fi
\ifdim\paperwidth=0pt\relax\else\pdfpagewidth\paperwidth\fi
\fi\fi\fi

\begin{document}
\textbf{This is the peer reviewed version of the following article: \\
\\
Aalen OO et al. Time-dependent mediators in survival analysis: Modeling direct and indirect effects with the additive hazards mode. Biometrical Journal. 2020; 62(3):532-549
which has been published in final form at [DOI: 10.1002/bimj.201800263].\\
\\
This article may be used for non-commercial purposes in accordance with Wiley Terms and Conditions for Use of Self-Archived Versions.}

\title{Time-dependent mediators in survival analysis: Modelling direct and indirect
effects with the additive hazards model}

\author[1]{Odd O. Aalen }
\author[2]{Mats J. Stensrud}
\author[3,4]{Vanessa Didelez}
\author[5]{Rhian Daniel}
\author[1]{Kjetil R{\o}ysland}
\author[6]{Susanne Strohmaier}

\affil[1]{Oslo Center for Biostatistics and Epidemiology, Department for Biostatistics, IMB, University of Oslo, Oslo, Norway}
\affil[2]{Department of Medicine, Diakonhjemmet Hospital, Oslo, Norway}
\affil[3]{Leibniz Institute for Prevention Research and Epidemiology - BIPS, Bremen, Germany}
\affil[4]{Faculty of Mathematics / Computer Science,
University of Bremen, Bremen, Germany}
\affil[5]{Division of Population Medicine, Cardiff University, UK}
\affil[6]{Institute of Clinical Biometrics, Medical University of Vienna, Vienna, Austria }

\date{Fall 2018}
\maketitle


\begin{abstract}
We discuss causal mediation analyses for survival data and propose a new approach based on the additive hazards model. The emphasis is on a dynamic point of view, that is, understanding how the direct and indirect effects develop over time. Hence, importantly, we allow for a time varying mediator. To define direct and indirect effects in such a longitudinal survival setting we take an interventional approach \citep{didelez18} where treatment is separated into one aspect affecting the mediator and a different aspect affecting survival. In general, this leads to a version of the non-parametric g-formula \citep{robins1986new}. In the present paper, we demonstrate that combining the g-formula with the additive hazards model and a sequential linear model for the mediator process results in simple and interpretable expressions for direct and indirect effects in terms of relative survival as well as cumulative hazards. Our results generalise and formalise the method of dynamic path analysis \citep{fosen06b,strohmaier15}. An application to data from a clinical trial on blood pressure medication is given.
\end{abstract}

\maketitle                   






\section{Introduction}
Mediation analysis has become a popular topic in causal inference, where it is used, broadly speaking, to understand the mechanism(s) through which a particular exposure affects an outcome. It attempts to distinguish between the direct and indirect effect, the latter being the portion of the total effect that passes through a certain variable, believed possibly to lie on a causal pathway from exposure to outcome, known as the mediator. The book by \cite{vanderweele15} contains an excellent overview of these ideas. The natural direct and indirect effects play an important role in this theory.

We are going to consider mediation in the context of survival analysis. 
An early paper estimating natural direct and indirect effects with the additive hazards model was 
presented by \cite{lange11}; for related papers, see \cite{martinussen11,nguyen2016,huang2017causal}.
Here we extend the approach to studying the effects of
mediators measured at a number of times throughout the observation period,
estimating their cumulative effects as a function of time. However, in this more general setting we need to 
go beyond the classical context of natural direct and indirect effects.

Natural direct and indirect effects were introduced by \cite{robins92}; see also \cite{pearl2001}. Let $A$ denote the exposure with two possible values,
$a$ and $a^{\ast}$. Let $M$ denote a mediator with the two corresponding potential outcomes
$M(a)$ and $M(a^{\ast})$. Let $Y$
denote the outcome of interest with $Y(a,m)$ denoting the potential outcome when the treatment is set to
$a$ and the mediator is set to $m$. Following \cite{vanderweele15} the natural direct effect ($NDE$) and natural
indirect effect ($NIE$) are defined as%

\begin{align*}
NDE=  &  E(Y(a,M(a^{*}))-E(Y(a^{*},M(a^{*})), \\
NIE=  &  E(Y(a,M(a))-E(Y(a,M(a^{*})).
\end{align*}
Note that this definition requires that we conceive of manipulating the mediator for each exposed individual to what would have occurred under non-exposure, and this manipulation cannot be performed in any hypothetical experiment. Adding up the direct and the indirect effect yields $E(Y(a,M(a))-Y(a^{*},M(a^{*}))$ which is the total effect.

When considering mediation in survival analysis difficulties can arise. 
 Consider two counterfactual
scenarios for an individual, the exposed and the non-exposed, and a mediator
which is measured at some time $t>0$ (given survival). One is then confronted
with the issue that an individual may survive up to time $t$ in one scenario
and not in the other, or he may be censored in one scenario and not in the
other. This means that the manipulation of the mediator required for natural effects may not make sense, and hence these effects may be ill-defined in a survival context.

There is a weaker concept which is useful in this context, namely that of
randomized interventional analogues of natural direct and indirect effects,
see e.g. \cite{vanderweele15}. The idea was introduced by \cite{geneletti07}, developed further by \cite{didelez06} and
applied to survival analysis by \cite{zheng2012} and by
\cite{lin2017mediation}. Instead of imagining manipulating the mediator for
each exposed individual to what would have occurred under non-exposure, one
draws a mediator value randomly from the non-exposed group conditional on confounders. Under no post-treatment and no unobserved mediator-outcome confounding the randomized intervention approach has the same identifiying formula as for the natural effects, but in principle it is a different target of inference; see also \cite{lok2016}. When it does not coincide with the natural effects, then the corresponding randomized interventional notions of direct and indirect effect do not add up to the usual total effect \citep{didelez06}.

Here we shall use a third approach which was developed by \cite{didelez18} based on 
work by \cite{robins2010alternative}. This is not based on manipulating the mediator, 
but on a different notion of mediation which is particularly suitable for survival analysis. 
One assumes the existence of variables that denote separate aspects of the treatment $A$: a
variable $A^M$ that influences the outcome through the mediator, and a variable  $A^D$ 
that has an effect directly on the outcome. The method is described in more detail in 
Section \ref{treatsep}. For a nonsurvival outcome $Y$,  $NDE$ and $NIE$ correspond in spirit to $E(Y|\mbox{do}(A^D=a, A^M=a^{*})) - E(Y|\mbox{do}(A^D=a^{*}, A^M=a^{*}))$ and $E(Y|\mbox{do}(A^D=a, A^M=a)) - E(Y|\mbox{do}(A^D=a, A^M=a^{*}))$ but are conceptually different as the point of intervention is a different one. One key advantage is that the causal assumptions are relatively straightforward and can be more easily understood. Moreover, the approach requires researchers to think more carefully about what are the separate aspects of treatment that constitute the direct and indirect pathways.

The aim of the paper is to show how the approach by \cite{didelez18} can be applied in survival analysis.
Our tool will be the additive hazards model which has seen an increasing use in causal inference
\citep{martinussen2010dynamic,martinussen11,martinussen13,tchetgen14}. A
useful feature of this model is that it can be analyzed just like any linear
model, which is a particular advantage in mediation analysis. It is also flexible and 
can be generalized to a dynamic
setting where direct and indirect effects are estimated as functions of time,
and where the mediators themselves may be time-varying. Clearly, this will be
of interest since for a process developing in time it would be sensible to
allow for mediation to be a time-dependent feature; see \citet[Section~9.2]{aalen08}.
Most often, mediation is presented in terms of time-fixed direct and indirect effects, while
it is more natural to view mediation as a dynamic concept. The aim of mediation analysis
is to understand mechanisms, and these are best understood in a
time-varying context.

A time-continuous stochastic process view of mediation is studied in
\cite{aalen2016can,aalen2018feedback} and it is shown how misleading results may follow when the
time aspect is not properly taken care of. Although we here only have time-discrete
mediator measurements available, we are still able to represent key aspects of of the time-varying structure.

The method we present here is an extension of dynamic path analysis which was developed by
\cite{fosen06b}; see also \cite{roysland2011analyzing} and \cite{aalen08}.
Dynamic path analysis was originally not defined in a formal causal inference
setting; it was first put on a causal footing by \cite{strohmaier15} who give
several results and a medical application. This paper is a continuation of the
work of \cite{strohmaier15} in a more formal causal inference framework. 
By using the approach of \cite{didelez18} we also avoid philosophical problems with counterfactuals and survival, like the issue of manipulating the mediator independently of the treatment given. 

The idea of dynamic path analysis is as follows. At each event time a standard
linear path analysis is carried out based on the treatment, the observed
mediator values and the observed event (i.e. the observed jump in a counting
process). The estimated direct and indirect effects are then integrated over
time to produce cumulative effect estimates. Although the estimates at each
event time are not in themselves useful, the cumulative estimates are well
defined and informative since we use the additive hazards model. We have
a sequence of linear models each giving mediation estimates valid for a small
time interval, and then these effects are added up over time. We shall show
that this is a meaningful procedure in a causal inference setting.

We wish to emphasize that the additive hazards 
model allows estimation of the parameters for each event time, and the 
parameters can vary arbitrarily over time. This yields a high degree of 
flexibility which is exploited in the present approach.

\section{Causal modeling of survival and mediators}

\label{causmed}

\begin{figure}[ptb]
\centering

\includegraphics[width=0.65\textwidth]{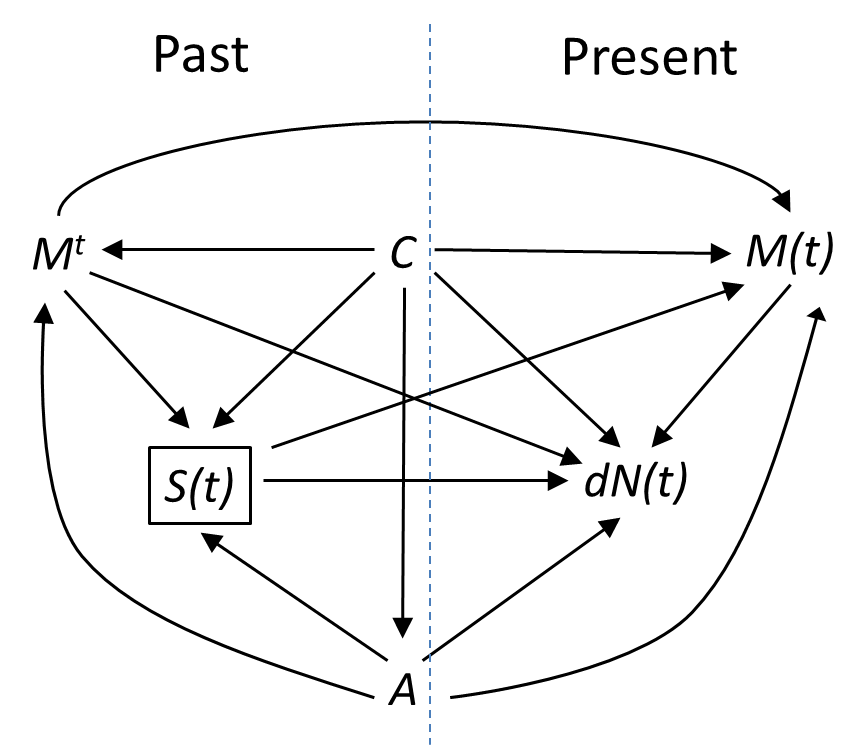}\caption{A diagram
illustrating the relationships between treatment $A$, mediator $M(t)$, set of
previous mediator values $M^{t}$, baseline covariates $C$, and outcome $dN(t)$ for any
time interval $(t,t+dt)$. The node $S(t)$ denotes survival up to time $t$; the
box around this node indicates that conditioning on survival is taking place. } %
\label{dyna13a}%
\end{figure}

\subsection{Basic concepts}

We consider a counting process $N(t)$ with a single possible event (standard
survival setting). The focus is on the
outcome $dN(t)$ for a small time interval $[t,t+dt)$. The treatment, or exposure,
is denoted $A$ and is given at baseline. The time of event is denoted $T$.  Covariates are given by
the vector $C$ which is measured at baseline. These covariates
are external (exogenous), that is, not influenced by the treatment $A$ nor by the development of 
the mediator processes. Baseline mediator values may be part of the covariates. 
The assumed relationship between the treatment,
the outcome $dN(t)$, and other quantities is shown in the 
diagram in Figure \ref{dyna13a}, which is defined at any time $t$. (More extensive diagrams showing the detailed development over time can also be drawn. These give important additional information; see \cite{didelez18}.) As seen from the figure
we consider a (scalar) mediator process $M(t)$ which
contains measurements for individuals who are still alive at time $t$.
Further, $M^{t}$ denotes the past mediator history, i.e. the set of all
previous mediator values. The mediator $M(t)$
must be known just prior to time $t$ in order to satisfy the requirements of
counting process theory. In practice, one might for instance use the last
measurement of the mediator prior to time $t$. Often the mediators will be
values from a stochastic process.

We assume
that the treatment, or exposure, of an individual is fixed at
time zero. Hence, treatment will not be changed according to changes in
covariates or mediators. Furthermore, we make the independent censoring assumption conditional on $C$ \citep{aalen08}.

The node $S(t)$ in Figure \ref{dyna13a} indicates whether the individual has
survived up to time $t$. Hence, $S(t)=1$ corresponds to $T \geq t$.

In our context, the additive hazard model takes the following form:

\begin{equation}
\lambda(t|M^{t}=m^{t}, A=a,C=c)  =\mu_t+\alpha
_{t}a+\beta_{t,g}^{\prime}m^t+\rho_{t}^{\prime}c %
\label{additive-general}
\end{equation}
with parameters $\mu_t$, $\alpha_{t}$, $\beta_{t,g}$, $\rho_{t}$. The subscript $g$ of $\beta_{t,g}$ means "general" and is used to distinguishing this model from the more limited one in equation (\ref{addhaz}).

\subsection{A treatment separation approach}
\label{treatsep}

\begin{figure}[ptb]
\centering
\includegraphics[width=0.65\textwidth]{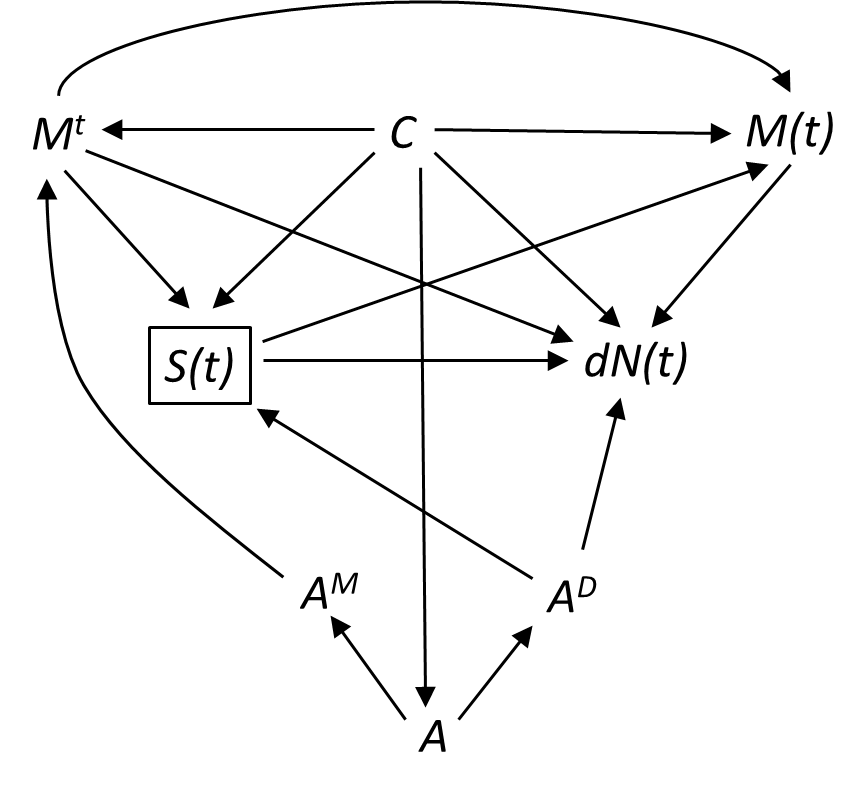}\caption{An augmented 
version of the graph in Figure \ref{dyna13a}. Following \cite{robins2010alternative} 
and \cite{didelez18} it is shown how different aspects of the treatment is 
separated into different values, $A^{M}$ and $A^D$, for the treatment effect on the 
mediators and the direct treatment effect on the outcome.}%
\label{dyna13f}%
\end{figure}

\begin{figure}[ptb]
\centering
\includegraphics[width=1.0\textwidth]{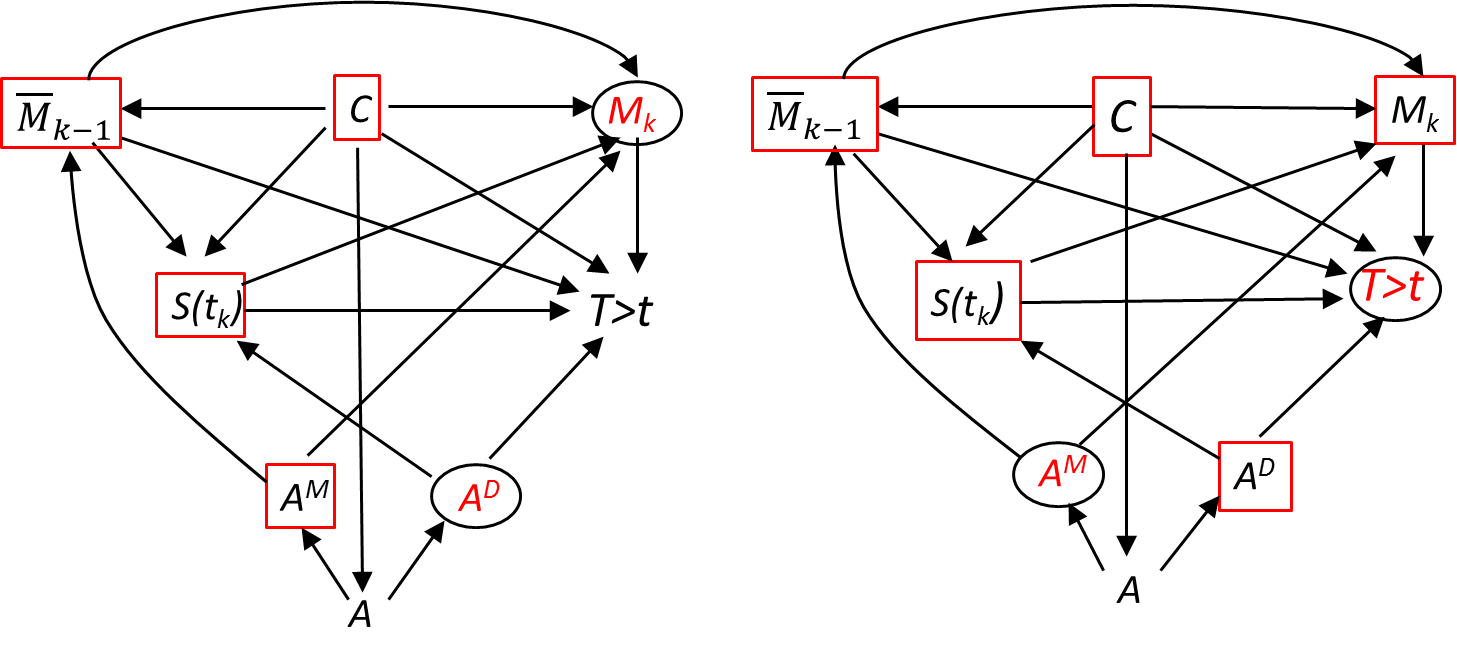}\caption{Using a time-discrete version of Figure \ref{dyna13f} to
demonstrate assumptions A1 (left panel) and A2 (right panel). In the left panel the two variables that are assumed conditionally independent are shown in red within circles, and the variables that one conditions on in assumption A1 are shown in boxes. The right panel gives a similar representation for assumption A2. By following the paths in the diagrams one sees that assumptions A1 and A2 are fulfilled. }%
\label{dyna13g}%
\end{figure}

We shall follow the approach developed by \cite{didelez18} based on work by \cite{robins2010alternative}. This 
can be illustrated in an augmented graph; see Figure \ref{dyna13f} where $A^M$ and $A^D$ denote separate aspects of the treatment $A$ that, respectively, influence the mediator, or have an effect directly on the survival outcome. Thus we imagine that the treatment has two separate biological components which can at least in principle (hypothetically) be manipulated separately. For instance, consider the statin treatment for prevention of heart disease. This will have an effect on risk of cardiovascular disease since statins reduce the cholesterol level; this is the indirect effect. However, there is also evidence that statins have other effects that influence disease risk \citep{stancu2001statins}; this would be the direct effect.

Note that $A^M$ and $A^D$ are variables in a hypothetical intervention situation, while $a$ and $a^{\ast}$ refer to possible values the treatment variables can take. In the observational setting, the values that $A$, $A^M$ and $A^D$ can take must all be the same, that is $A=A^M=A^D$ (all identical) . We assume Property 1 of \cite{didelez18}, saying that $P(T>t|\mbox{do} (A^D=a, A^M=a))= P(T>t|\mbox{do} (A=a))$. The total effect is a contrast of setting both $A^M=a$ and $A^D=a$ versus setting both $A^M=a^{\ast}$ and $A^D=a^{\ast}$. The direct effect is then defined by varying only the value of $A^D$ while keeping $A^M$ fixed, and vice versa for the indirect effect. Here $a$ is the value of the treatment, e.g. "$a$=no treatment" and "$a^{\ast}$=treatment".

The aim is to estimate the quantity $P(T>t|\mbox{do} (A^D=a, A^M=a^{\ast}))$, from which direct and indirect effects may be derived by varying the values of $A^D$ and $A^M$, respectively. For this purpose one can derive a mediational g-formula, a time-discrete version of which is given by \cite{didelez18}. \cite{lin2017mediation} derive the same formula but for a different target of inference based on the randomized intervention approach.

We wish to write a time-continuous version of the mediational g-formula. This can be derived from the g-computation formula for survival as shown below. However, while the survival model is time-continuous, the mediators are defined discretely in time. This is a limitation, but corresponds to the way measurements normally are taken in a survival study, where mediators are measured at given time points.

We define a discrete mediator process as follows: Let $t_{0}=0<t_{1}<\ldots<t_{k}<\ldots$ be an increasing sequence of time points and let $\{M_k,k=0, 1,\ldots\}$ be a sequence of mediator values. Define
the mediator process $M(t)$ as follows: $M(t)=M_k$ for $t_{k}\leq t<t_{k+1}$ for
$k=0,1,2,\ldots $. Let $\overline{M}_k$ denote the set of all $M_i$ for $i \leq k$, and define $r(t)=k$ when $t_{k}\leq t<t_{k+1}$. The mediators are only defined as long as the individual survives, that is $T > t$, otherwise they are undefined.

Modifying slightly the assumptions of \cite{didelez18} we make the following three assumptions for the validity of a mediational g-formula.

\begin{itemize}
\item {A0: The covariates $C$ are sufficient to adjust for confounding between $A$ and all $M_k$ as well as between $A$ and $T$. Remark: As seen from the diagram in Figure \ref{dyna13a}, $C$ must be set of baseline covariates, and conditioning on it will remove confounding. We can then use results from \cite{didelez18} which assumes a randomized treatment.}

\item {A1: For each time $t_k$ the mediator
$M_k$ is independent of the treatment component $A^D$  conditional on $T > t_k$, previous mediator values, $A^M$, and covariates $C$:

\hspace{0.2cm} $M_k \mathrel{\text{\scalebox{1.07}
{$\perp\mkern-10mu\perp$}}}A^D| (T > t_k, \overline{M}_{k-1},A^M=a^{\ast}, C)$}

\item {A2: For each $k$ and for each time $t$ between $t_k$ and $t_{k+1}$ the event $T > t$ is independent of the treatment component $A^M$  conditional on $T > t_k$, previous mediator values, $A^D$ and covariates $C$:

\hspace{0cm} $T > t
\mathrel{\text{\scalebox{1.07}{$\perp\mkern-10mu\perp$}}}A^M|( T > t_k, \overline{M}_k, A^D=a,C)$}
\end{itemize}

Conditions A1 and A2 imply that, conditional on $C$, there is no mediator-outcome confounding. However, these assumptions go beyond this. The intuitive meaning of the assumptions is that the paths corresponding to treatment components $A^M$ and $A^D$ are separated, such that the effect of $A^M$ passes through the mediator. In a biological context, this means that the effect via the mediator and the direct effect are separate biological phenomena, not intertwined with each other. So, one is making an explicit statement about biology (or some other substance matter), which is left implicit in other approaches to mediation analysis. For instance, when defining natural direct and indirect effects one assumes that one can imagine  intervening on the mediator after having made the treatment intervention. This approach clearly requires a biological basis to be taken seriously.

A demonstration of the validity of assumptions A1 and A2 for the setting studied in this paper is given in the diagrams of Figure \ref{dyna13g}. The variables that are assumed conditionally independent are shown in red in the two panels corresponding to assumption A1 and A2 respectively. The conditioning variables are shown in boxes. One can check the paths between the relevant variables to see that they are all closed, which implies conditional independence.

\subsection{A total probability formula}
\label{totpr}

The mediational g-formula is closely related to a simple law of total probability which we derive briefly below. This is also close to the standard g-computation formula; see \cite{lok2004estimating} for a continuous-time version. We want to stress the intuitive content of the formulas and the relation to the likelihood.

Consider survival over the time interval $(0,t)$. We condition on baseline covariates $C$. The observations are given as a sequence of measurements of mediators and survival between mediators. For $t_{k}\leq t<t_{k+1}$ the joint distribution of these quantities conditional on the treatment $A=a$ can be written as a product of conditional probabilities:

\begin{align*}
& P(T>t,\overline{M}_{k}=\overline{m}_{k}|A=a,C) \\
& = P(T>t, T>t_1, T>t_2, \ldots,T>t_k,\overline{M}_{k}=\overline{m}_{k}|A=a,C) \\
& =P(T > t|T > t_{k},\overline{M}_{k}=\overline{m}_{k}, A=a,C) \\
& \times \prod_{t_i \leq t} \Big\{P(T > t_i|T > t_{i-1},\overline{M}_{i-1}=\overline{m}_{i-1}, A=a,C) \\
& \times P(M_i=m_i|T > t_i,\overline{M}_{i-1}=\overline{m}_{i-1}, A=a,C) \Big\}. 
\end{align*}
Following \cite{lok2004estimating} we define variables that are indexed by -1 as not present. Hence, e.g. $P(M_i=m_i|T > t_i,\overline{M}_{i-1}=\overline{m}_{i-1}, A=a,C)$ is to be read as $P(M_0=m_0|A=a,C)$ when $i=0$.

By integrating out the mediators we get  the marginal survival probability given $A=a$ and $C$. (Here we write this as a sum, but for continuous mediators this would be substituted by integrals.)

\begin{align*}
& P(T>t|A=a,C) \\
& = \sum_{ \overline{m}_k} \Big[ P(T > t|T > t_{k},\overline{M}_{k}=\overline{m}_{k}, A=a,C) \\
& \times \prod_{t_i \leq t} \big\{ P(T > t_i|T > t_{i-1},\overline{M}_{i-1}=\overline{m}_{i-1}, A=a,C)  \\
& \times   P(M_i=m_i|T > t_i,\overline{M}_{i-1}=\overline{m}_{i-1}, A=a,C) \big\}  \Big].  
\end{align*}

From the no unmeasured confounding assumption A0 it follows that $A=a$ can be substituted by $\mbox{do}(A=a)$ on the left hand side. This gives the classical g-computation formula \citep{lok2004estimating}:

\begin{align}
& P(T>t|\mbox{do}(A=a),C) \nonumber \\
& = \sum_{ \overline{m}_k} \Big[ P(T > t|T > t_{k},\overline{M}_{k}=\overline{m}_{k}, A=a,C) \nonumber \\
& \times \prod_{t_i \leq t} \big\{ P(T > t_i|T > t_{i-1},\overline{M}_{i-1}=\overline{m}_{i-1}, A=a,C)  \label{totprob} \\
& \times   P(M_i=m_i|T > t_i,\overline{M}_{i-1}=\overline{m}_{i-1}, A=a,C)\big\}  \Big] \nonumber .
\end{align}

\subsection{A mediational g-formula}

Assume assumptions A0, A1 and A2 are fulfilled. Using the approach of \cite{didelez18} we can substitute $A=a$ by separate interventions in formula (\ref{totprob}); $A=a$ in the first and second line and $A=a^{\ast}$ in the third line. This gives the mediational g-formula:

\begin{align*}
& P(T>t|\mbox{do} (A^D=a, A^M=a^{\ast}),C) \\
& = \sum_{ \overline{m}_k} \Big[ P(T > t|T > t_{k},\overline{M}_{k}=\overline{m}_{k}, A=a,C) \\
& \times \prod_{t_i \leq t} \big\{ P(T > t_i|T > t_{i-1},\overline{M}_{i-1}=\overline{m}_{i-1}, A=a,C)  \\
& \times   P(M_i=m_i|T > t_i,\overline{M}_{i-1}=\overline{m}_{i-1}, A=a^{\ast},C)\big\}  \Big]  .
\end{align*}
Note that our approach is not about intervening on the mediator, so it is not the classical g-formula in that sense, although mathematically the formula is the same.

Let the observed hazard rate, or intensity process \citep{aalen08}, be defined by $\lambda(t|\overline{M}_{r(t)}=\overline{m}_{r(t)}, A=a,C)$. Then we can write for $t_{i}\leq t<t_{i+1}$:

\begin{equation*}
P(T > t|T > t_{i},\overline{M}_{i}=\overline{m}_{i}, A=a,C) =  \exp \big\{-\int_{t_{i}}^t \lambda(s|\overline{M}_{i}=\overline{m}_{i}, A=a,C) ds  \big\}
\end{equation*}

Hence, the mediational g-formula can be written as:

\begin{equation}
\begin{split}
& P(T>t|\mbox{do} (A^D=a, A^M=a^{\ast}),C) \\
&  =  \sum_{\overline{m}_k} \Big[ \exp \big\{-\int_0^t \lambda(s|\overline{M}_{r(s)}=\overline{m}_{r(s)}, A=a,C) ds  \big\}  \label{medg}\\ 
&\times  \prod_{t_i \leq t}  P(M_i=m_i|T > t_i,\overline{M}_{i-1}=\overline{m}_{i-1}, A=a^{\ast},C) \Big].    
\end{split}
\end{equation}

Since we have now rephrased the mediational g-formula in terms of hazard functions the latter can be estimated under censoring as usual, and we assume independent censoring as stated previously.

Formula (\ref{medg}) is generally applicable to hazard rate models. However, we get particularly simple solutions when applying an additive hazards model. Hence, we leave the general setting and make specific assumptions. First we shall define a model for the mediators.

\section{Special case: Models for the mediators}

\label{simplifying model}

\subsection{Two linear models}

We shall assume that the mediators follow a sequential linear model. This has several advantages, e.g. giving a simple preservation of the structure when conditioning on survival under an additive hazards model. A similar linear model was used by \cite{strohmaier15} (in their Section 3 and in Proposition 1 of their Appendix).

For $i=0,\ldots,n$ assume the following mediator model: For those who survive at time $t_i$ the mediator value is given by

\begin{equation}
M_{i} = \lambda_{i} A +\delta _{i}^{\prime}C +\sum_{k<i} b_{ik} M_{k}+\varepsilon_{i}.
\label{structural}
\end{equation}
(For $i=0$ the sum $\sum_{k<i} b_{ik} M_{k}$ in the above expression is equal to zero). For those who do not survive at time $t_i$ the mediator $M_i$ is undefined. Here $A$ is the treatment and $C$ is the, possibly multidimensional, covariate. Given survival at time $t_i$ the remainder term $\varepsilon_{i}$ is assumed to be independent of all $\varepsilon_{k}$ for $k<i$, and also independent of $A$ and $C$. The $\lambda$'s, $\delta$'s and $b$'s are coefficients.

In order to apply the mediational g-formula we shall also need a marginal  model for each $M_i$ as a solution to the linear system (\ref{structural}). Conditionally on $T>t_i$, assume that we can write:

\begin{equation}
M_i=m_{0,i}+\gamma_i A+\theta_i^{\prime}C+\eta_i
\label{huff901}%
\end{equation}
for suitable parameter values  $m_{0,i}$, $\gamma_i$ and $\theta_i$, and where the remainder term $\eta_i$ is independent of $A$ and $C$. 
Model (\ref{huff901}) can be derived from model (\ref{structural}) under a certain assumption; see Section \ref{cons}. The independence of the remainder term in (\ref{huff901}) implies that estimation of $\gamma_i$ and $\theta_i$ can be done by regressing $M_i$ on $A$ and $C$.

\subsection{Consistency of models}
\label{cons}

We have introduced the two linear models (\ref{structural}) and (\ref{huff901}) for the mediators. We shall show that by  introducing an additive hazards assumption the latter model can be derived from the first one.

Let $\overline{\varepsilon}_i$ be defined similarly to $\overline{M}_i$ as the vector of $\varepsilon_{k}$ for $k \leq i$, and assume $T > t_i$. In matrix form model (\ref{structural}) can be written as follows with the solution with respect to           $\overline{M}_i$ on the right:

\begin{equation}\overline{M}_i =A \Lambda_i  + \Delta_i C +B_i \overline{M}_i +\overline{\varepsilon}_i  ,\;\text{\quad \quad }\;\;\overline{M}_i =(I -B_i)^{ -1} (A \Lambda_i  + \Delta_i C +\overline{\varepsilon}_i). \label{solution1}
\end{equation}
Here, $\overline{M}_i$, $\Lambda_i$, and $\overline{\varepsilon}_i$ are column vectors, $\Delta_i$ is a matrix, and $B_i$ is a strictly lower triangular matrix consisting of the components $b_{ik}$.

From the right-hand side of equation (\ref{solution1}) one sees that $\overline{M}_{i}$ is a linear 
combination of $A$, $C$ and $\overline{\varepsilon}_{i}$. Hence, corresponding to equation (\ref{huff901}), $M_i$ can be written in the form (conditional on $T>t_i$):

\begin{equation}
M_i = m_{0,i} + \gamma_i A + \theta_i^{\prime}C + \sum_{k\leq i} g_{i,k} \varepsilon_k
\label{betasum}
\end{equation}
for coefficients $g_{i,k}$. In order to derive the representation (\ref{huff901}), we need to prove that the last term (the sum) in (\ref{betasum}) is independent of $A$ and $C$ given $T>t_i$. In fact we shall also prove that all $\varepsilon_k$ for $k \leq i$ are independent of each other as well as of $A$ and $C$ given $T>t_i$. We shall assume an additive hazards model; that is, the hazard rate  $\lambda(t|\overline{M}_{r(t)}=\overline{m}_{r(t)}, A=a,C)$ is a linear combination of $A$, $C$ and the mediators. We shall use an argument from \citet[Section~3.1]{strohmaier15} about preservation of independent covariates under survival, given the additive model. However, we have to be careful because the mediator $M_i$ is only defined for those surviving up to time $t_i$ and does not exist before this time. Hence we have to argue in a stepwise fashion.

We first need a slight generalization of the argument in \citet[Section~3.1]{strohmaier15}. Define $V=(V_1,V_2)$ and $X_1, X_2, \ldots, X_k$ as covariates at baseline in a hazard model. Assume that the variables $V, X_1, X_2, \ldots, X_k$ are all independent, but that $V_1$ and $V_2$ may be mutually dependent. Furthermore, assume that we have an additive hazards model with these variables as covariates. Then the variables $V, X_1, X_2, \ldots, X_k$ are still independent given survival up to time $t$. The proof is given in the Lemma in the Appendix.

Assume $t_{k}\leq t<t_{k+1}$, then by (\ref{structural}) and (\ref{solution1}) the hazard rate $\lambda(t|\overline{M}_{r(t)}=\overline{m}_{r(t)}, A=a,C)$ is also a linear combination of  $A$, $C$ and $\overline{\varepsilon}_{k}$. 

We shall use an argument by induction and start by setting $k=0$. Hence we start at time $t_0=0$ and consider the interval up to time $t_1$. The covariates in this first interval are given at baseline as $A$, $C$ and $M_0$. The hazard rate is a linear combination of these quantities, or alternatively of  $A$, $C$ and $\varepsilon_0$. By model (\ref{structural}) $(A,C)$ and $\varepsilon_0$ are independent quantities. However, note that the components of $A$ and $C$ may be mutually dependent; see Figure \ref{dyna13a}. From the Appendix it follows that the independence of $(A,C)$ and $\varepsilon_0$ is preserved conditional on survival up to time $t_1$. At this time covariate $M_1$ is introduced by adding the independent quantity     $\varepsilon_1$; see equation (\ref{structural}).

This implies that for those who survive up to $t_1$ there will be three independent quantities at this time, namely  $(A,C)$, $\varepsilon_0$ and $\varepsilon_1$. Using the argument from the Appendix once more, it follows that these three quantities are still independent given survival up to the next time, $t_2$. At this time, another independent quantity, $\varepsilon_2$ is added.

Continuing this stepwise argument it follows that conditional on survival up to $t_i$ the bivariate quantity $(A,C)$, and the components of $\overline{\varepsilon}_{i}$ will all be independent. This implies the validity of formula (\ref{huff901}) where $\eta_i$ is independent of $(A,C)$ given $T > t_i$. We conclude that indeed, under the additive hazards assumption, equation (\ref{huff901}) follows from equation (\ref{structural}).

\section{Mediation analysis under the additive hazards model and the linear model for the mediators}

\label{stochint}

\subsection{A special case of the additive hazards model}

We shall use the following additive hazards model as a special case of model (\ref{additive-general}):
\begin{equation}
\lambda(t|\overline{M}_{r(t)}=\overline{m}_{r(t)}, A=a,C=c)  =\mu_t +\alpha_{t}a+\beta_{t}m_{r(t)}+\rho_{t}^{\prime}c %
\label{addhaz}
\end{equation}
where $m_{r(t)}$ is the value of the mediator at time $t$. Note that by definition the hazard rate, or intensity process, is defined dependent on $T>t$. In formula (\ref{addhaz}) we make the simple assumption that only the last value of the mediator has an impact. This assumption corresponds to removing the arrow from $M^t$ to $dN(t)$ in Figures \ref{dyna13a} and \ref{dyna13f}. Then it is seen that   the mediator value $M(t)$ is sufficient to model the hazard function. When this quantity is included, the remaining parts of the mediation process do not have any further influence at time $t$.

\subsection{Applying the mediational g-formula}

We shall now apply the mediational g-formula to the particular models for the mediators and the hazard rate. This will result in simple formulas which allow explicit expressions for the direct and indirect effects, mirroring the results of dynamic path analysis \citep{strohmaier15,fosen06b}.

We start by introducing the hazard rate in equation (\ref{addhaz}) into the mediational g-formula given in equation (\ref{medg}). Non-random quantities are taken outside the summation sign:

\begin{align*}
& P(T>t|\mbox{do} (A^D=a, A^M=a^{\ast}),C=c) \\
&  =   \sum_{\overline{m}_k} \Big[ \exp \big\{-\int_0^t (\mu_s+\alpha_{s}a+\beta_{s}{m}_{r(s)}+\rho_{s}^{\prime}c)ds) \big\}   \\
&\times  \prod_{t_i \leq t}  P(M_i=m_i|T > t_i,\overline{M}_{i-1}=\overline{m}_{i-1}, A=a^{\ast},C=c) \Big]
\\
&  =  \exp \big\{-\int_0^t (\mu_s+\alpha_{s}a+\rho_{s}^{\prime}c)ds) \big\} 
 \sum_{\overline{m}_k} \Big[ \exp \big\{-\int_0^t (\beta_{s}{m}_{r(s)} ds) \big\}   \\
&\times  \prod_{t_i \leq t}  P(M_i=m_i|T > t_i,\overline{M}_{i-1}=\overline{m}_{i-1}, A=a^{\ast},C=c) \Big] .     
\end{align*}

For the quantity $M_{r(s)}$ we use the decomposition in formula (\ref{huff901}) and again take non-random quantities outside the summation sign:

\begin{align}
& P(T>t|\mbox{do} (A^D=a, A^M=a^{\ast}),c) \nonumber \\
&  =  \exp \big\{-\int_0^t (\mu_s+\alpha_{s}a+\rho_{s}^{\prime}c)ds) \big\} 
  \exp \big\{-\int_0^t (\beta_{s}(m_{0,r(s)}+\gamma_{r(s)} a^{\ast}+\theta_{r(s)}^{\prime}c)ds) \big\} \nonumber\\
& \times \sum_{\overline{m}_{k}} \Big[ \exp \big\{-\int_0^t (\beta_{s}{\eta_{r(s)}} ds) \big\}  \nonumber \\   
& \times   \prod_{t_i < t}  P(M_{i}=m_i |T > t_i,\overline{M}_{i-1}=\overline{m}_{i-1}, A=a^{\ast},C=c) \Big]  \label{bigsum1}.
\end{align}

By formula  (\ref{structural}) we can write:

\begin{align*}
& P(M_{i}=m_i |T > t_i,\overline{M}_{i-1}=\overline{m}_{i-1}, A=a^{\ast},C=c) \\
& = P(A \lambda_{i}+\delta _{i}^{\prime}C +\sum_{k<i} b_{ik} M_{k}+\varepsilon_{i}=m_i |T > t_i,\overline{M}_{i-1}=\overline{m}_{i-1}, A=a^{\ast},C=c) \\
& = P(\varepsilon_{i}=m_i - a^{\ast} \lambda_{i}-\delta _{i}^{\prime}c -\sum_{k<i} b_{ik} m_{k} |T > t_i,\overline{M}_{i-1}=\overline{m}_{i-1}, A=a^{\ast},C=c) \\
& = P(\varepsilon_{i}=e_i |T > t_i,\overline{M}_{i-1}=\overline{m}_{i-1}, A=a^{\ast},C=c) \\
& = P(\varepsilon_{i}=e_i |T > t_i) .
\end{align*}

Given $A$ and $C$ there is a 1-1 relationship
between the $\overline{\varepsilon}_{k}$ and the $\overline{M}_k$; see equation (\ref{structural}). Hence, we can just as well sum over all values  $\overline{\varepsilon}_{k}=\overline{e}_k$ when handling the sum in equation (\ref{bigsum1}): 

\begin{align}
& P(T>t|\mbox{do} (A^D=a, A^M=a^{\ast}),c) \nonumber \\
&  =  \exp \big\{-\int_0^t (\mu_s+\alpha_{s}a+\rho_{s}^{\prime}c)ds) \big\} 
  \exp \big\{-\int_0^t (\beta_{s}(m_{0,r(s)}+\gamma_{r(s)} a^{\ast}+\theta_{r(s)}^{\prime}c)ds) \big\} \nonumber\\
& \times \sum_{\overline{e}_{k}} \Big[ \exp \big\{-\int_0^t (\beta_{s}{\eta_{r(s)}} ds) \big\}   
 \times   \prod_{t_i < t}  P(\varepsilon_{i}=e_i |T > t_i) \Big]  \label{bigsum}
\end{align}
where $\eta_{r(s)}$ (i.e. conditional with respect to $T>t_{r(s)}$) is a linear combination of the $\varepsilon_{k}$ by the last term in equation (\ref{betasum}) and independent of $A$ and $C$ by the result in Section \ref{cons}. Hence the sum given by line (\ref{bigsum}) is independent of $a^{\ast}$ and $c$.

Hence, we can write the mediational g-formula in the following way:

\begin{align}
& Q(t;a,a^{\ast},c)=P(T>t|\mbox{do} (A^D=a, A^M=a^{\ast}),c) \nonumber \\
& =  f(t,c) \exp \big\{- a \int_0^t \alpha_{s}ds -a^{\ast} \int_0^t \beta_s \gamma_{r(s)} ds \big\}
\label{bigmed}
\end{align}
where $f(t,c)$ is functionally independent of $a$ and $a^{\ast}$.

Taking the logarithm of $P(T>t|\mbox{do} (A^D=a, A^M=a^{\ast}),c)$ and changing sign yields the cumulative hazard rate. Differentiating with respect to $t$ gives the corresponding hazard rate $h(t|\mbox{do} (A^D=a, A^M=a^{\ast}),c)$:

\begin{equation}
q(t;a,a^{\ast},c)=h(t|\mbox{do} (A^D=a, A^M=a^{\ast}),c) 
=\alpha_{t}a+\beta_{t}\gamma_{r(t)}a^{\ast} -\frac{\delta}{\delta t}\log(f(t,c))
\end{equation}
This is a mediational g-formula for the hazard rate.

\subsection{Direct and indirect effects}

Direct and indirect effects can be defined both in terms of the survival functions \citep{didelez18} or as contrasts of the hazard rates. Considering first the survival function, we can define the survival indirect and direct effects in a relative manner, and we denote these as $SIE$ and $SDE$ respectively. We define

\begin{align*}
SIE(t)  &  = Q(t;a,a,c)/Q(t;a,a^{\ast},c)=\exp  \big\{ (a^{\ast}-a) \int_0^t  \beta_{s} \gamma_{r(s)}ds \big\}   \\
SDE(t)  &  = Q(t;a,a^{\ast},c)/Q(t;a^{\ast},a^{\ast},c)= \exp \big\{ (a^{\ast}-a) \int_0^t\alpha_{s}ds \big\} 
\end{align*}
One sees that these quantities are independent of $c$, hence they are the same as the direct and
indirect effects marginalized over the covariate $C$.

$SIE$ describes the multiplicative factor for the survival probability if $a$ is applied along the $A^M$ path instead of $a^*$; it can be seen to depend cumulatively on $\beta_s$, the coefficients of $m_{r(s)}$ in the hazard model (\ref{addhaz}), and $\gamma_{i}$, the coefficient of A in model (\ref{huff901}) which has re-expressed $M_i$ in terms of all past error terms.

We can also define indirect and direct effects based on differences between  survival functions instead of dividing them as we did above. In this case the function $f(t,c)$ does not disappear, but can be estimated by using the total effects.

Alternatively, the above survival direct and indirect effects correspond to the following hazard rate differences:

\begin{align*}
HIE(t,c)  &  =(q(t;a,a,c)-q(t;a,a^{\ast},c))=(a-a^{\ast})\beta_{t}%
\gamma_{r(t)}\\
HDE(t,c)  &  =(q(t;a,a^{\ast},c)-q(t;a^{\ast},a^{\ast},c))=(a-a^{\ast
})\alpha_{t}
\end{align*}
One sees that these quantities are independent of $c$, hence they also represent direct and
indirect effects marginalized over the covariate $C$. Notice that these formulas can be seen as time-local versions of the product formulas of \cite{baron1986moderator}.

In practice, we would use cumulative versions of indirect and direct effects
integrated over time as follows:%

\begin{equation}
CHIE(t)=(a-a^{\ast})\int_{0}^{t}\gamma_{r(s)}\beta_{s}ds,\quad CHDE%
(t)=(a-a^{\ast})\int_{0}^{t}\alpha_{s}ds
\label{cdir}%
\end{equation}
This corresponds to results given in \citep{fosen06b,aalen08,strohmaier15}. A
proof for these as natural indirect and direct effects is given
in Corollary 1 in the online eAppendix of \cite{lange11} for the special case
when the mediator is fixed at time 0, while in our case the mediator changes
over time.

The total effect is given as:

\begin{equation}
CHTE(t)=CHIE(t)+CHDE(t) = (a-a^{\ast})\int_{0}^{t}(\gamma_{r(s)}\beta_{s}+\alpha_{s})ds
\label{total}
\end{equation}

Notice that the simple expressions above are specific to the assumptions we have here, especially that we have a linear model for mediators and an additive hazards model, both without interactions. Using only the last mediator value also gives a simplification by allowing us to use model (\ref{huff901}). Nevertheless the models are useful in practice as demonstrated in the example given here as well as in \cite{strohmaier15}. We conjecture that more complex models can be used, but would result in more complex formulas or require Monte Carlo methods. It would be an interesting topic for further research to formulate a general approach as an alternative to the "natural effect" models of \cite{lange11}.

\section{Estimation in dynamic path analysis}

\label{dynpa}

We now describe the actual estimation of the above causal parameters. This follows very much the methods of \citep{fosen06b,strohmaier15} and essentially relies on applying linear models at each event time.
Even though
$dN(t)$ is defined for jumps in a counting process, the results on the classical 
linear models, including least squares
estimation, are valid. This analogy is used throughout the theory of the
additive hazards model, and follows from counting process theory
\citep{aalen08}. Clearly, due to the infinitesimal nature of the time interval
these effects would not be of much use in themselves. However, we shall
integrate these effects over time to obtain interpretable cumulative estimates.
This is analogous to the cumulative hazard rates and the cumulative
regression functions in the additive hazards model \citep{aalen08}. The
time-local effects will correspond to derivatives of these curves.

\subsection{Estimation of cumulative direct and indirect effects}

\label{estimation}

Consider the cumulative direct and indirect effects defined by the formulas in (\ref{cdir}).
For estimation purposes we consider $n$ independent copies of the variables
defined above, that is counting processes $N^{(j)}(t)$, mediators $M^{(j)}_i$, exposures $A^{(j)}$ and covariates $C^{(j)}$ for $j=1,...,n$. By
counting process theory least squares estimation in the additive hazards model
gives essentially unbiased estimators; see \citet[page~158]{aalen08}. Estimation in the
regression equation for the mediator is done by standard least squares
analysis at each event time. Details about the estimation may be found in \citep{fosen06b,aalen08,strohmaier15}. Statistical properties like consistency and asymptotic normality can be derived from general results for the additive hazards model and stochastic integrals for counting processes. The theoretical background is given in \citep{aalen08}. We only give a summary presentation here:

\begin{enumerate}
\item The regression functions $\mu_t$, $\alpha_{t}$, $\beta_{t}$, $\rho_{t}$ in
equation (\ref{addhaz})
are estimated by the additive hazards regression model, i.e. in a cumulative
fashion. The resulting cumulative estimates are denoted $\hat{\mu}_{0}(t)$,
$\hat{A}(t)$, $\hat{B}(t)$ and $\hat{R}(t)$.

\item For all $i$ the parameters $\gamma_{i}$ and $\theta_i$ in equation (\ref{huff901}) are estimated  by standard linear regression of $M_i$ on $A$ and $C$ for those who survive up to time $t_i$. Denote
these estimates $\hat{\gamma}_i$ and $\hat{\theta}_i$. (Note that these estimates deviate somewhat from the standard approach in dynamic path analysis, where the mediator values are regressed on $A$ and $C$ at each event time. The numerical difference appears to be small in practice.)

\item The integral $\int_{0}^{t}\gamma_{r(s)}\beta_{s}ds$ is estimated by
multiplying each increment in the estimate of $\int_{0}^{t}\beta_{s} ds$ by
the relevant estimate of $\gamma_{r(s)}$. From equation (\ref{cdir}), the indirect effect of $A$ on
the hazard rate is estimated by the cumulative regression function%
\[
\widehat{CHIE}(t)=(a-a^{*})\int_{0}^{t}\hat{\gamma}_{r(s)} d\hat{B}(s)
\]
and the direct effect is estimated by
\[
\widehat{CHDE}(t)=(a-a^{*})\widehat{A}(t).
\]

\item  Direct and indirect effects for the survival function can be estimated by taking the exponential function of minus the quantities estimated here.

\item Standard errors can be computed using the bootstrap.
\end{enumerate}

\subsection{Interpretation of direct and indirect effects}

The interpretation of the cumulative effects $CHIE(t)$ and $CHDE(t)$ is similar to that of the
cumulative regression effects in the additive hazards model. Local effects at
given times can be understood as slopes of the cumulative curves.

One important aspect of our approach is to illuminate how mediation may change
over time. Consider as an example a clinical trial comparing two cholesterol
treatments with the outcome being a major coronary event as in
\cite{strohmaier15}. One may for instance be interested in whether the
treatment effect mediated by cholesterol is changing over time; indeed, it appears from the curves in \cite{strohmaier15} that an
increasing proportion of effect is mediated over time.

Note that when expressing effects in terms of survival functions through $SIE(t)$ and $SDE(t)$, we are not dependent on interpreting the hazard rate. This is desirable in causal inference settings, see \cite{ryalen2018transforming}.

\subsection{Adjusting for measurement error in the mediator}
\label{measurerror}

It is well known that uncertainty in the measurement of the mediator may lead to bias in the estimation of direct and indirect effects. Since we have a linear structure we can use the formulas in \cite[Section~3.5.1]{vanderweele15}. In our context this is a tentative approach which should be studied in more detail, but it makes good sense due to the formal similarities (linear regression, least squares analysis) between our setting and that of VanderWeele. Here, it is assumed that there is a mis-measured mediator $\widetilde{M}=M+\varepsilon$ where $\varepsilon$ is normally distributed with zero mean and independent of $M$. Let $\kappa$ denote the proportion of variance in $\widetilde{M}$ explained by $M$.

Assume $\widetilde{\alpha}_s$ and  $\widetilde{\beta}_s$ are estimated from the procedure in bullet point 1 in Section \ref{estimation} using $\widetilde{M}$ instead of $M$. Following \cite[fomulas~(3.5)]{vanderweele15} we get the following modified values for the quantities in equation (\ref{cdir}).

\begin{equation}
\beta_{s}=\widetilde{\beta}_s / \kappa, \qquad
\alpha_{s}= \widetilde{\alpha}_s - (\widetilde{\beta}_s \gamma_{r(s)}) / \kappa + \widetilde{\beta}_s \gamma_{r(s)}, 
\label{correc}%
\end{equation}

The total effect is given by formula (\ref{total}):

\begin{equation*}
CHTE(t)= (a-a^{\ast})\int_{0}^{t}(\gamma_{r(s)}\beta_{s}+\alpha_{s})ds = (a-a^{\ast})\int_{0}^{t}(\gamma_{r(s)}\widetilde{\beta}_{s}+\widetilde{\alpha})ds
\end{equation*}

Hence, the total effect is not affected by measurement error in the mediator, as is to be expected. From formula (\ref{correc}) it follows that the adjustment due to measurement error in  the mediator is simply to divide the estimated indirect effect $\widehat{CHIE}(t)$ by $\kappa$ and then subtract this from the estimated total effect  to get the estimated $\widehat{CHDE}(t)$. Confidence intervals may be estimated by using the bootstrap.

\section{Analysis of the SPRINT data}
\subsection{The SPRINT study in brief}
We obtained access to individual level data from the Systolic Blood Pressure Intervention Trial \citep{sprint2015randomized} through the SPRINT data challenge. The SPRINT study was a randomized, controlled, open-label, multi-center trial undertaken in the United States. More details regarding the trial conduct can be found in the original New England Journal of Medicine article \citep{sprint2015randomized} and the supplementary material mentioned therein.

In brief, more than 9000 participants without diabetes, but with systolic blood pressure (SBP) levels of 130 mm Hg or higher and increased cardiovascular risk, were randomized to either intensive or standard blood pressure treatment. 
In patients in the intensive treatment group, the systolic blood pressure target value was 120 mmHg, compared to the target value of 140 mmHg in the standard treatment strategy group. This corresponds to a binary treatment indicator $A$ taking either value 'intensive' or 'standard'. We consider this to be a fixed binary treatment variable since the patients are assigned to one of the treatment regimes at time $t=0$. In practice however, achieving the target systolic blood pressure values required adjusting dosage and combination of different blood pressure medications. Particularly, blood pressure values were monitored on a monthly basis for the first 3 months and every 3 months thereafter, when participants returned to the study site. Based on current and previous systolic blood pressure values medications were adjusted to achieve the respective target values. The primary outcome was a composite endpoint including myocardial infarction, acute coronary syndrome not resulting in myocardial infarction, stroke, acute decompensated heart failure, or death from cardiovascular causes. After a median follow-up of 3.26 years, a significant beneficial effect of the intensive treatment on the composite cardiovascular outcome was reported.

\subsection{Applying dynamic path analysis to analyse the relationship between the SPRINT intervention, diastolic blood pressure and acute kidney injury or failure}
Despite the beneficial effect of the intervention on the primary  outcome, a significantly increased risk for certain serious adverse events was reported for the intensive treatment group (Table 3 in \cite{sprint2015randomized}), particularly for acute kidney injury or failure. It has been suggested that under intensive therapy, blood pressure may be more likely to fall below the  threshold for autoregulation of kidney perfusion \citep{rocco2018effects}. Hence, we decided to employ dynamic path analysis to study how much of the effect on kidney injury/failure  (AKI) was mediated through diastolic blood pressure. In this simple illustrative example, we have simply censored subjects at the time they experience competing events, such as death, but we acknowledge that competing events should treated more carefully in future applications.

We assume that the blood pressure intervention has two components that could be manipulated separately:  One component $A^M$ affects diastolic blood pressure, and another component $A^D$ affects kidney failure through other pathways. This is not an innocuous assumption, as it cannot be directly tested. In particular, the component influencing diastolic blood pressure must be distinct from the component influencing systolic blood pressure. Today there is no antihypertensive agent that only influences diastolic blood pressure, or only influences systolic blood pressure. However, antihypertensive agents have differential effects on diastolic blood pressure and systolic blood pressure, which may suggest that different pathways influence diastolic blood pressure and systolic blood pressure \citep{wu2005summary}. Also, it seems plausible that blood pressure agents may cause kidney failure independently of their effect on diastolic blood pressure. In particular, there is well-established physiological evidence that some antihypertensive agents (used in the SPRINT trial), are associated with higher risk of acute kidney failure than other agents, even though they have similar effects on blood pressure \citep{rose2001clinical}. 
Hence, we aim to study a hypothetical intervention in which the treatment component affecting diastolic blood pressure $A^M$ is set to standard treatment.

Note that the treatment decomposition is not about being “testable” or not (like normality or linearity) - in some situations it may be possible / can be imaged that treatment can be decomposed as required, in others not - in the former our target of inference corresponds to a meaningful real-world quantity, in the latter not.

We restricted the analysis to 9342 participants who had valid diastolic blood pressure measurements at each of their up to 21 recorded measurement times. Of these, 4670 were assigned to the intensive group and 4673 to the standard group. Overall, 324 AKI events were observed (204 in the intensive treatment group, 120 in the standard treatment group). Randomisation of the intervention should ensure that assumption A0 is satisfied. We considered age, sex, race (black/non black), smoking status (never, past, current), body mass index ($kg/m^2$, cont.), serum creatinine ($mg/dl$, cont.), high density lipoprotein cholesterol ($mg/dl$ , cont.), number of blood pressure agents before randomisation as well as clinical and subclinical cardiovascular disease as potential confounding variables for the mediator-outcome relationship aiming to satisfy assumptions A1 and A2. To facilitate causal interpretation, we present both additive hazard estimates and relative survival curves. We acknowledge that for some practical purposes it would be more desirable to provide survival curves on the difference scale, however, it would entail estimation of $f(t,c)$ e.g. in equation (\ref{bigmed}), which will be studied in future research.

\begin{figure}
	\centering
  \includegraphics[width=1\textwidth]{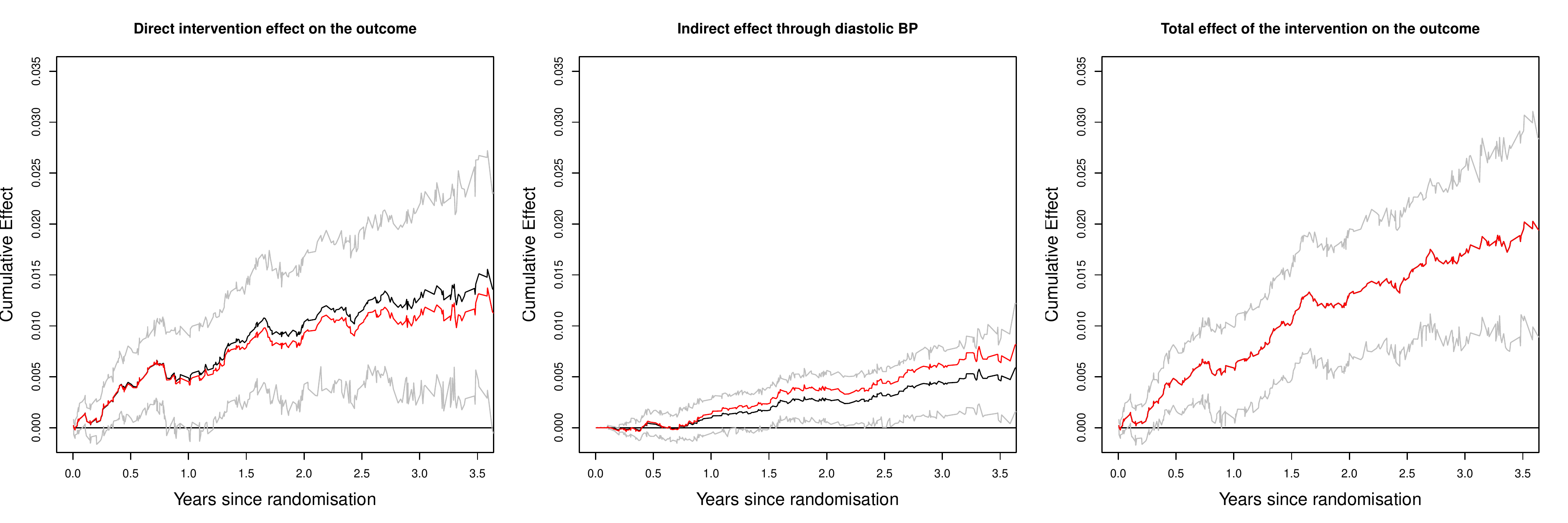}
	\caption{Estimated direct ( $\widehat{CHDE}(t)$), indirect ( $\widehat{CHIE}(t)$) and total effect ( $\widehat{CHTE}(t)$) on the\textbf{ cumulative hazard scale} comparing the effects of intensive to standard blood pressure treatment on the risk of acute kidney injury/failure (AKI) mediated through diastolic blood pressure (DBP). Black lines show effect estimates, gray lines the corresponding 95\% confidence intervals and red lines represent effect estimates corrected for measurement error.}
	\label{fig4}
\end{figure}

\begin{figure}
	\centering
  \includegraphics[width=0.7\textwidth]{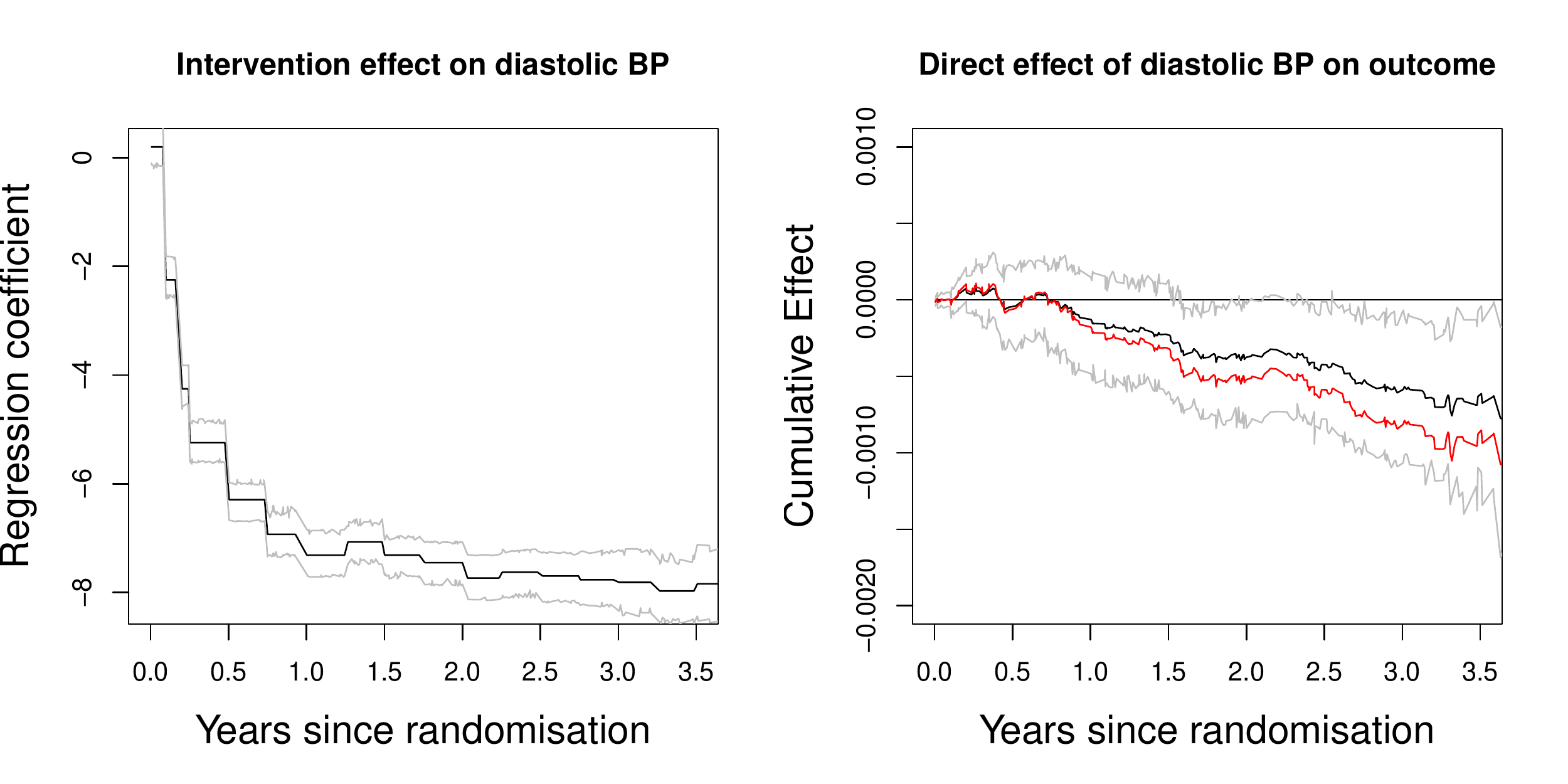}
	\caption{Estimated regression coefficients over time for the effect of intensive compared to standard blood pressure treatment on diastolic blood pressure (DBP) (i.e. estimated parameter $\gamma_i$ from formula (\ref{huff901})), and direct effect of DBP on AKI on the \textbf{cumulative hazard scale} (i.e. estimated function $\hat{B}(t)$ from Section \ref{estimation}).  Black lines show effect estimates, gray lines the corresponding 95\% confidence intervals and red lines represent effect estimates corrected for measurement error.}
	\label{fig5}
\end{figure}

\begin{figure}
	\centering
  \includegraphics[width=1\textwidth]{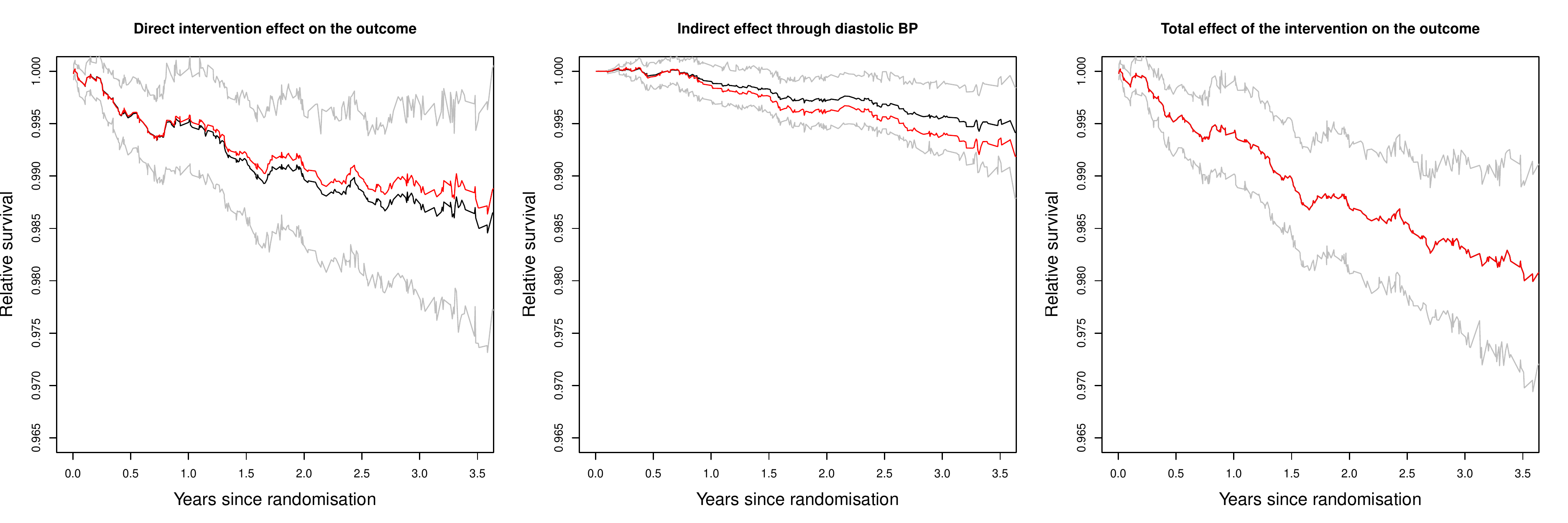}
	\caption{Estimated direct ( $\widehat{SDE}(t)$), indirect ( $\widehat{SIE}(t)$) and total effect ( $\widehat{STE}(t)$) on the \textbf{relative survival scale} comparing the effects of intensive to standard blood pressure treatment on the risk of acute kidney injury/failure (AKI) mediated through diastolic blood pressure (DBP). Black lines show effect estimates, gray lines the corresponding 95\% confidence intervals and red lines represent effect estimates corrected for measurement error.}
	\label{fig6}
\end{figure}

\begin{figure}
	\centering
  \includegraphics[width=0.7\textwidth]{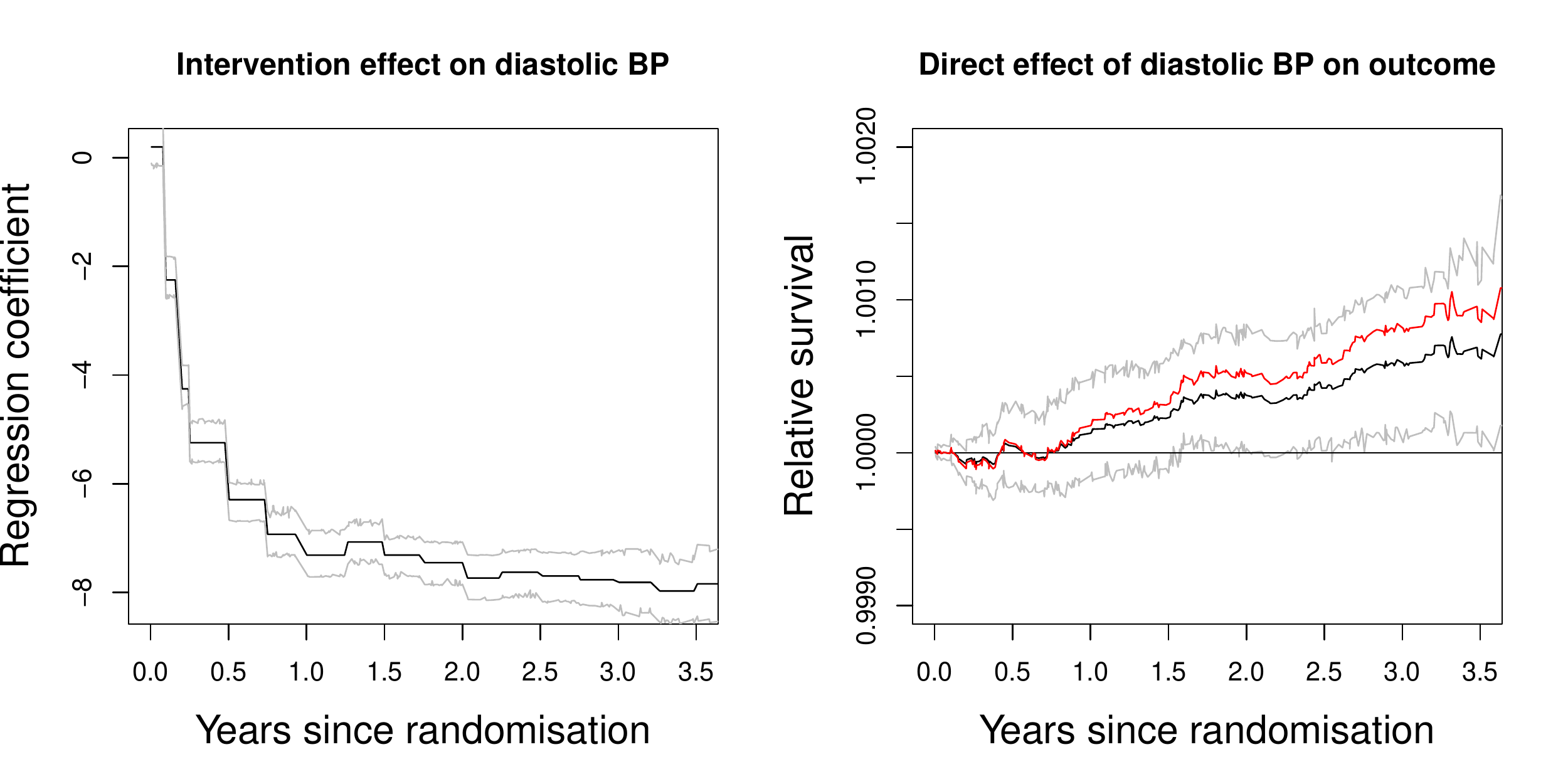}
	\caption{Estimated regression coefficients over time for the effect of intensive blood pressure compared to standard treatment on diastolic blood pressure (DBP) (i.e. estimated parameter $\gamma_i$ from formula (\ref{huff901})), and direct effect of DBP on AKI on the \textbf{relative survival scale} (i.e. estimated function $\exp(-\hat{B}(t))$, see Section \ref{estimation}). Black lines show effect estimates, gray lines the corresponding 95\% confidence intervals and red lines represent effect estimates corrected for measurement error.}
	\label{fig7}
\end{figure}

The results of our analyses are presented in Figures \ref{fig4} to \ref{fig7} on different scales. Figure \ref{fig4} shows $\widehat{CHDE}(t)$, $\widehat{CHIE}(t)$ and $\widehat{CHTE}(t)$ of the intensive compared to standard treatment strategy. The total effect was obtained by summing the direct and indirect effect. However, we also compared these results with the total effect obtained from an outcome model completely ignoring the mediator and got the same results. Figure \ref{fig5} shows the linear regression coefficients of the intervention on DBP, as well as the direct effect of DBP on the AKI outcome on the cumulative hazard scale. Figures \ref{fig6} and \ref{fig7} show the effects on the relative survival scale (corresponding to the direct effect $SDE(t)$, the indirect effect $SIE(t)$ and the total effect $STE(t)=SDE(t)\times SIE(t)$). 
The black lines display effect estimates calculated from the observed mediator values. The gray lines are confidence intervals based on 200 bootstrap samples. The red line represents the corrected effect estimates, using the correction approach described in Section \ref{measurerror}, where $\kappa$ was set to $0.72$ based on results by \cite{intraclass2016}. Confidence intervals for the corrected estimates can also be obtained by bootstrapping. However, to keep the figure parsimonious, we have omitted these curves.

As reported in \cite{sprint2015randomized}, an increased risk for acute kidney injury or failure can be observed for the intensive treatment group. We find that only  a small, but significant, effect is mediated through diastolic blood pressure. That is, modifying the intensive treatment such that the diastolic blood pressure is similar to standard treatment will only have a minor impact on the hazard of kidney failure. On the other hand, the direct effect is substantial: intensive treatment seems to increase the risk of kidney failure mostly through other pathways than reducing diastolic blood pressure. To uncover these pathways, other mediators may be considered in future work. We suspect that one such pathway may involve reduction of the systolic blood pressure, which is the actual aim of intensive blood pressure treatment. Therefore, the increase in risk of acute kidney failure may be an inherent side effect of the treatment that is not preventable. 

Overall, on both the cumulative hazard and relative survival scale the effects appear fairly linear over time. It only takes approximately half a year until the diastolic blood pressure values reach a stable low level, which may indirectly influence the AKI risk. For this particular example,  we could have obtained similar results using constant effects in the additive hazard model. However, it is only by using an approach that allows for the effects to vary over time that it was possible to reveal that in fact the effects appear close to constant. Furthermore, it can be observed that measurement error in the mediator can lead to overestimation of the direct effect and underestimation of the indirect effect.

\subsection{Program code}
In our online supporting  material we provide R code to perform dynamic path analysis on simulated data that mimick the main features of the real life application.

\section{Discussion}

The proposed approach to mediation analysis with longitudinal mediators and survival outcomes provides a formal causal footing and extends existing approaches to the problem.
The key features of our approach are that we phrase the target of inference in terms of separate aspects of treatment while assuming an additive hazards model and a piecewise constant mediator process that follows a linear model. These modelling choices result in appealingly simple explicit solutions which are relatively straightforward to estimate.

Note that as detailed in  \citep{didelez18}  (and as has been noted by other authors), natural direct and indirect effects are not really defined in a survival setting with longitudinal mediator (if observing the mediator stops at death). \cite{lange11} define natural direct and indirect effects for a special case of our setting, i.e. for a mediator fixed at time 0, and they are then identical to the effects we define here. In a more general setting, natural effects may be substituted by the randomized interventional analogues; see \cite{lin2017mediation}.

In the present work we have made a number of simplifying assumptions which future work should aim to address:
we assume a piecewise constant mediator process, corresponding to the time-discrete way in which the mediators are observed. This may induce a slight misspecification of the mediator model. Indeed, it may of course be a continuous underlying mediator process, which however is not observable. Our approach corresponds to the standard procedure in survival analysis where time-dependent covariates are updated at the times they are observed. Modelling the underlying processes would be far more complex and is usually not attempted. Nevertheless, this may imply some violation of assumption A2. In further work a sensitivity analysis of this issue should be performed, e.g. generalising our measurement error model. Clearly, the importance of this problem depends on the frequency of mediator measurements. In clinical trials, patients are usually recorded at frequent visits, and the problem may be less severe.

Further, we assume a relatively simple mediator model, where the hazard depends only on the last observed mediator value.
This can be modified and mediators can for instance be defined as linear functions of the past. The issue of multiple mediators is also an important theme and we believe the theory can be extended to cover this along the lines of \cite{vansteelandt2017interventional}. Another simplifying assumption of our  hazard model is the absence of any exposure-mediator interactions; we believe that results may likely be extended to allow interactions along the lines of \cite{vanderweele15}. Furthermore, the case of a fixed treatment as well as confounding just by baseline covariates can likely be relaxed by generalising the approach of \cite{didelez18}.

Time-dependent confounding is an important issue, but it is beyond the setting we have here. Time dependent confounding can in principle be dealt with, if it is not itself directly affected by treatment \citep{didelez18}.

A feature of all survival analysis is that an unobserved frailty variable may complicate the interpretation of hazard rates; see e.g. \cite{aalen2015}. We also intend to follow this up along the lines of \cite{stensrud2017exploring} and \cite{valberg2018surprising}. Note that when focusing on survival quantities like $SIE(t)$ and $SDE(t)$ this is not an issue, see \cite{ryalen2018transforming}.

The additive hazards model may occasionally yield non-monotone estimates of cumulative hazard functions or survival functions. This may be an indication that the additive model is not valid; see
\cite{aalen08} for ways to check the model. If the deviation is relatively minor, one can handle such monotonicity issues by simple ad hoc approaches, see \citet[page~64-65]{lin1994semiparametric}, \citet[page~174]{aalen08} and \cite{huang2017restoration}. Using the Cox model instead would result in a more complex mediational g-formula and, likely, would not result in explicit solutions for the direct and indirect effects in terms of survival probabilities. While we believe that the additive hazard model has many advantages and is very flexible, it would be interesting to carefully compare our approach with possible alternatives such as the Cox model for mediation problems in future work.

\section*{Acknowledgements}
This manuscript was prepared using SPRINT POP Research Materials obtained form the NHLBI Biologic Specimen and Data Repository Information Coordinating Center and does not necessary reflect the opinions or views of the SPRINT POP or the NHLBI.

Susanne Strohmaier has received funding from the European Union’s Horizon 2020 research and innovation programme under the Marie Sk\l odowska-Curie grant agreement No 795292. Rhian Daniel is supported by a Sir Henry Dale Fellowship jointly funded by the Wellcome Trust and the Royal Society (Grant 107617/Z/15/Z). Mats Julius Stensrud and Kjetil R\o ysland were supported by the Research Council of Norway, grant NFR239956/F20 - Analyzing clinical health registries: Improved software and mathematics of identifiability.


\vspace*{1pc}

\noindent {\bf{Conflict of Interest}}

\noindent {\it{The authors have declared no conflict of interest. }}

\section*{Appendix}

We shall generalize slightly the argument in \citet[Section~3.1]{strohmaier15}.

{\it Lemma}: Define $V=(V_1,V_2)$ and $X_1, X_2, \ldots, X_k$ as covariates at baseline in a survival model. Assume that the variables $V, X_1, X_2, \ldots, X_k$ are all independent, but that the components $V_1$ and $V_2$ may be mutually dependent. Assume that we have an additive hazards model with $V_1,V_2, X_1, X_2, \ldots, X_k$ as covariates. We shall show that the variables $V, X_1, X_2, \ldots, X_k$ are still independent given survival up to time $t$.

{\it Proof}: With an additive hazards model (with no interaction) the probability of surviving up to time $t$ is of the form.

\begin{equation}
\theta(t)\exp(- a_1(t) V_1 - a_2(t) V_2  -b_1(t) X_1 -\ldots -b_k(t)X_k)
\label{cumcum}
\end{equation}

Conditional on survival the joint probabilities of the covariates is found as follows:

\begin{align*}
& P(V =v,X_1=x_1, \ldots, X_k=x_k |T>t)\\
&  =\frac{P(V =v,X_1=x_1, \ldots, X_k=x_k ,T>t)}{P(T>t)}\\
&  =\frac{P(T>t|V =v,X_1=x_1, \ldots, X_k=x_k )f_V(v) f_{X_1}(x_1) \times \ldots \times f_{X_k}(x_k)}{P(T>t)}%
\end{align*}
where the $f_u(.)$ denotes covariate distributions at time 0. Putting in the survival probability (\ref{cumcum}) gives us:%

\begin{align*}
& P(V =v,X_1=x_1, \ldots, X_k=x_k |T>t)\\
&  =(1/P(T>t)) \{\theta(t)\exp(- a_1(t) v_1 - a_2(t) v_2  -b_1(t) x_1 -\ldots -b_k(t)x_k) \\
& \times f_V(v) f_{X_1}(x_1) \times \ldots \times f_{X_k}(x_k)\}%
\end{align*}
The conditional probability distribution can be factorized and hence $V, X_1, X_2, \ldots, X_k$ are still independent at time $t$ conditional on survival.

\bibliographystyle{wileyauy}

\bibliography{papers6}

\begin{thebibliography}{39}
\providecommand{\natexlab}[1]{#1}
\providecommand{\url}[1]{\texttt{#1}}
\providecommand{\urlprefix}{URL }
\expandafter\ifx\csname urlstyle\endcsname\relax
  \providecommand{\doi}[1]{doi:\discretionary{}{}{}#1}\else
  \providecommand{\doi}{doi:\discretionary{}{}{}\begingroup
  \urlstyle{rm}\Url}\fi
\input{babelbst.tex}
\newcommand{\Capitalize}[1]{\uppercase{#1}}
\newcommand{\capitalize}[1]{\expandafter\Capitalize#1}
\providecommand\eprint[2][]{{\if!#1!#2\else\href{http://www.#1.org/#2}{#2}\fi}}

\bibitem[{Aalen \emph{\bbletal{}}(2016)Aalen, R{\o}ysland, Gran, Kouyos,
  \bbland{} Lange}]{aalen2016can}
Aalen, O., R{\o}ysland, K., Gran, J., Kouyos, R., \bbland{} Lange, T. (2016)
  Can we believe the dags? a comment on the relationship between causal dags
  and mechanisms. \emph{Statistical methods in medical research},
  \textbf{25}~(5), 2294--2314.

\bibitem[{Aalen \emph{\bbletal{}}(2008)Aalen, Borgan, \bbland{}
  Gjessing}]{aalen08}
Aalen, O.O., Borgan, {\O}., \bbland{} Gjessing, H.K. (2008) \emph{Survival and
  event history analysis: a process point of view}, Springer.

\bibitem[{Aalen \emph{\bbletal{}}(2015)Aalen, Cook, \bbland{}
  R{\o}ysland}]{aalen2015}
Aalen, O.O., Cook, R.J., \bbland{} R{\o}ysland, K. (2015) Does cox analysis of
  a randomized survival study yield a causal treatment effect? \emph{Lifetime
  data analysis}, \textbf{21}~(4), 579--593.

\bibitem[{Aalen \emph{\bbletal{}}(2018)Aalen, Gran, R{\o}ysland, Stensrud,
  \bbland{} Strohmaier}]{aalen2018feedback}
Aalen, O.O., Gran, J.M., R{\o}ysland, K., Stensrud, M.J., \bbland{} Strohmaier,
  S. (2018) Feedback and mediation in causal inference illustrated by
  stochastic process models. \emph{Scandinavian Journal of Statistics},
  \textbf{45}~(1), 62--86.

\bibitem[{Baron \bbland{} Kenny(1986)}]{baron1986moderator}
Baron, R.M. \bbland{} Kenny, D.A. (1986) The moderator--mediator variable
  distinction in social psychological research: Conceptual, strategic, and
  statistical considerations. \emph{Journal of personality and social
  psychology}, \textbf{51}~(6), 1173.

\bibitem[{Didelez(2018)}]{didelez18}
Didelez, V. (2018) Defining causal meditation with a longitudinal mediator and
  a survival outcome. \emph{Lifetime Data Analysis},
  \doi{10.1007/s10985-018-9449-0}.

\bibitem[{Didelez \emph{\bbletal{}}(2006)Didelez, Dawid, \bbland{}
  Geneletti}]{didelez06}
Didelez, V., Dawid, P., \bbland{} Geneletti, S. (2006) Direct and indirect
  effects of sequential treatments, \bblin{} \emph{Proceedings of the
  Twenty-Second Conference Annual Conference on Uncertainty in Artificial
  Intelligence (UAI-06)}, AUAI Press, Arlington, Virginia, \bblpp{} 138--146.

\bibitem[{Filipovsky \emph{\bbletal{}}(2016)Filipovsky, Seidlerova, Kratochvil,
  Karnosova, Hronova, \bbland{} Mayer}]{intraclass2016}
Filipovsky, J., Seidlerova, J., Kratochvil, Z., Karnosova, P., Hronova, M.,
  \bbland{} Mayer, J.O. (2016) Automated compared to manual office blood
  pressure and to home blood pressure in hypertensive patients. \emph{Blood
  Pressure}, \textbf{25}~(4), 228--234.

\bibitem[{Fosen \emph{\bbletal{}}(2006)Fosen, Ferkingstad, Borgan, \bbland{}
  Aalen}]{fosen06b}
Fosen, J., Ferkingstad, E., Borgan, {\O}., \bbland{} Aalen, O.O. (2006) Dynamic
  path analysis -- a new approach to analyzing time-dependent covariates.
  \emph{Lifetime data analysis}, \textbf{12}~(2), 143--167.

\bibitem[{Geneletti(2007)}]{geneletti07}
Geneletti, S. (2007) Identifying direct and indirect effects in a
  non-counterfactual framework. \emph{Journal of the Royal Statistical Society:
  Series B (Statistical Methodology)}, \textbf{69}~(2), 199--215.

\bibitem[{Huang(2017)}]{huang2017restoration}
Huang, Y. (2017) Restoration of monotonicity respecting in dynamic regression.
  \emph{Journal of the American Statistical Association}, \bblpp{} 1--10.

\bibitem[{Huang \bbland{} Yang(2017)}]{huang2017causal}
Huang, Y.T. \bbland{} Yang, H.I. (2017) Causal mediation analysis of survival
  outcome with multiple mediators. \emph{Epidemiology}, \textbf{28}~(3),
  370--378.

\bibitem[{Lange \bbland{} Hansen(2011)}]{lange11}
Lange, T. \bbland{} Hansen, J.V. (2011) Direct and indirect effects in a
  survival context. \emph{Epidemiology}, \textbf{22}~(4), 575--581.

\bibitem[{Lin \bbland{} Ying(1994)}]{lin1994semiparametric}
Lin, D. \bbland{} Ying, Z. (1994) Semiparametric analysis of the additive risk
  model. \emph{Biometrika}, \textbf{81}~(1), 61--71.

\bibitem[{Lin \emph{\bbletal{}}(2017)Lin, Young, Logan, \bbland{}
  VanderWeele}]{lin2017mediation}
Lin, S.H., Young, J.G., Logan, R., \bbland{} VanderWeele, T.J. (2017) Mediation
  analysis for a survival outcome with time-varying exposures, mediators, and
  confounders. \emph{Statistics in medicine}, \textbf{36}~(26), 4153--4166.

\bibitem[{Lok \emph{\bbletal{}}(2004)Lok, Gill, Van Der~Vaart, \bbland{}
  Robins}]{lok2004estimating}
Lok, J., Gill, R., Van Der~Vaart, A., \bbland{} Robins, J. (2004) Estimating
  the causal effect of a time-varying treatment on time-to-event using
  structural nested failure time models. \emph{Statistica Neerlandica},
  \textbf{58}~(3), 271--295.

\bibitem[{Lok(2016)}]{lok2016}
Lok, J.J. (2016) Defining and estimating causal direct and indirect effects
  when setting the mediator to specific values is not feasible.
  \emph{Statistics in medicine}, \textbf{35}~(22), 4008--4020.

\bibitem[{Martinussen(2010)}]{martinussen2010dynamic}
Martinussen, T. (2010) Dynamic path analysis for event time data: large sample
  properties and inference. \emph{Lifetime data analysis}, \textbf{16}~(1),
  85--101.

\bibitem[{Martinussen \bbland{} Vansteelandt(2013)}]{martinussen13}
Martinussen, T. \bbland{} Vansteelandt, S. (2013) On collapsibility and
  confounding bias in cox and aalen regression models. \emph{Lifetime data
  analysis}, \textbf{19}~(3), 279--296, \doi{10.1007/s10985-013-9242-z}.

\bibitem[{Martinussen \emph{\bbletal{}}(2011)Martinussen, Vansteelandt,
  Gerster, \bbland{} Hjelmborg}]{martinussen11}
Martinussen, T., Vansteelandt, S., Gerster, M., \bbland{} Hjelmborg, J.v.B.
  (2011) Estimation of direct effects for survival data by using the aalen
  additive hazards model. \emph{Journal of the Royal Statistical Society:
  Series B (Statistical Methodology)}, \textbf{73}~(5), 773--788.

\bibitem[{Nguyen \emph{\bbletal{}}(2016)Nguyen, Tchetgen, Kawachi, Gilman,
  Walter, \bbland{} Glymour}]{nguyen2016}
Nguyen, T.T., Tchetgen, E.J.T., Kawachi, I., Gilman, S.E., Walter, S.,
  \bbland{} Glymour, M.M. (2016) Comparing alternative effect decomposition
  methods: The role of literacy in mediating educational effects on mortality.
  \emph{Epidemiology}, \textbf{27}~(5), 670--676.

\bibitem[{Pearl(2001)}]{pearl2001}
Pearl, J. (2001) Direct and indirect effects, \bblin{} \emph{Proceedings of the
  seventeenth conference on uncertainty in artificial intelligence}, Morgan
  Kaufmann Publishers Inc., \bblpp{} 411--420.

\bibitem[{Robins(1986)}]{robins1986new}
Robins, J. (1986) A new approach to causal inference in mortality studies with
  a sustained exposure period - application to control of the healthy worker
  survivor effect. \emph{Mathematical modelling}, \textbf{7}~(9-12),
  1393--1512.

\bibitem[{Robins \bbland{} Greenland(1992)}]{robins92}
Robins, J.M. \bbland{} Greenland, S. (1992) Identifiability and exchangeability
  for direct and indirect effects. \emph{Epidemiology}, \textbf{3}~(2),
  143--155.

\bibitem[{Robins \bbland{} Richardson(2011)}]{robins2010alternative}
Robins, J.M. \bbland{} Richardson, T.S. (2011) Alternative graphical causal
  models and the identification of direct effects. \emph{In P. Shrout, K. Keyes
  and K. Ornstein (Eds.), Causality and psychopathology: Finding the
  determinants of disorders and their cures}, \bblpp{} 103--158.

\bibitem[{Rocco \emph{\bbletal{}}(2018)Rocco, Sink, Lovato, Wolfgram, Wiegmann,
  Wall, Umanath, Rahbari-Oskoui, Porter, Pisoni
  \emph{\bbletal{}}}]{rocco2018effects}
Rocco, M.V., Sink, K.M., Lovato, L.C., Wolfgram, D.F., Wiegmann, T.B., Wall,
  B.M., Umanath, K., Rahbari-Oskoui, F., Porter, A.C., Pisoni, R.
  \emph{\bbletal{}} (2018) Effects of intensive blood pressure treatment on
  acute kidney injury events in the systolic blood pressure intervention trial
  (sprint). \emph{American Journal of Kidney Diseases}, \textbf{71}~(3),
  352--361.

\bibitem[{Rose \bbland{} Post(2001)}]{rose2001clinical}
Rose, B. \bbland{} Post, T. (2001) Clinical physiology of acid-base and
  electrolyte disorders (clinical physiology of acid base \& electrolyte
  disorders). \emph{5th Edition: McGraw-Hill Education}, \bblp{} 2001.

\bibitem[{R{\o}ysland \emph{\bbletal{}}(2011)R{\o}ysland, Gran, Ledergerber,
  Wyl, Young, \bbland{} Aalen}]{roysland2011analyzing}
R{\o}ysland, K., Gran, J.M., Ledergerber, B., Wyl, V., Young, J., \bbland{}
  Aalen, O.O. (2011) Analyzing direct and indirect effects of treatment using
  dynamic path analysis applied to data from the {S}wiss {H}{I}{V} {C}ohort
  {S}tudy. \emph{Statistics in medicine}, \textbf{30}~(24), 2947--2958.

\bibitem[{Ryalen \emph{\bbletal{}}(2018)Ryalen, Stensrud, \bbland{}
  R{\o}ysland}]{ryalen2018transforming}
Ryalen, P.C., Stensrud, M.J., \bbland{} R{\o}ysland, K. (2018) Transforming
  cumulative hazard estimates. \emph{Biometrika}, \textbf{105}~(4), 905--916.

\bibitem[{SPRINT(2015)}]{sprint2015randomized}
SPRINT (2015) A randomized trial of intensive versus standard blood-pressure
  control. \emph{N Engl J Med}, \textbf{2015}~(373), 2103--2116.

\bibitem[{Stancu \bbland{} Sima(2001)}]{stancu2001statins}
Stancu, C. \bbland{} Sima, A. (2001) Statins: mechanism of action and effects.
  \emph{Journal of cellular and molecular medicine}, \textbf{5}~(4), 378--387.

\bibitem[{Stensrud \emph{\bbletal{}}(2017)Stensrud, Valberg, R{\o}ysland,
  \bbland{} Aalen}]{stensrud2017exploring}
Stensrud, M.J., Valberg, M., R{\o}ysland, K., \bbland{} Aalen, O.O. (2017)
  Exploring selection bias by causal frailty models. \emph{Epidemiology},
  \textbf{28}~(3), 379--386.

\bibitem[{Strohmaier \emph{\bbletal{}}(2015)Strohmaier, R{\o}ysland, Hoff,
  Borgan, Pedersen, \bbland{} Aalen}]{strohmaier15}
Strohmaier, S., R{\o}ysland, K., Hoff, R., Borgan, {\O}., Pedersen, T.R.,
  \bbland{} Aalen, O.O. (2015) Dynamic path analysis -- a useful tool to
  investigate mediation processes in clinical survival trials. \emph{Statistics
  in medicine}, \textbf{34}~(29), 3866--3887.

\bibitem[{Tchetgen~Tchetgen \emph{\bbletal{}}(2015)Tchetgen~Tchetgen, Walter,
  Vansteelandt, Martinussen, \bbland{} Glymour}]{tchetgen14}
Tchetgen~Tchetgen, E.J., Walter, S., Vansteelandt, S., Martinussen, T.,
  \bbland{} Glymour, M. (2015) Instrumental variable estimation in a survival
  context. \emph{Epidemiology}, \textbf{26}~(3), 402--410.

\bibitem[{Valberg \emph{\bbletal{}}(2018)Valberg, Stensrud, \bbland{}
  Aalen}]{valberg2018surprising}
Valberg, M., Stensrud, M.J., \bbland{} Aalen, O.O. (2018) The surprising
  implications of familial association in disease risk. \emph{BMC public
  health}, \textbf{18}~(1), 135.

\bibitem[{VanderWeele(2015)}]{vanderweele15}
VanderWeele, T.J. (2015) \emph{Explanation in causal inference: methods for
  mediation and interaction}, Oxford University Press.

\bibitem[{Vansteelandt \bbland{} Daniel(2017)}]{vansteelandt2017interventional}
Vansteelandt, S. \bbland{} Daniel, R.M. (2017) Interventional effects for
  mediation analysis with multiple mediators. \emph{Epidemiology (Cambridge,
  Mass.)}, \textbf{28}~(2), 258--265.

\bibitem[{Wu \emph{\bbletal{}}(2005)Wu, Kraja, Oberman, Lewis, Ellison, Arnett,
  Heiss, Lalouel, Turner, Hunt \emph{\bbletal{}}}]{wu2005summary}
Wu, J., Kraja, A.T., Oberman, A., Lewis, C.E., Ellison, R.C., Arnett, D.K.,
  Heiss, G., Lalouel, J.M., Turner, S.T., Hunt, S.C. \emph{\bbletal{}} (2005) A
  summary of the effects of antihypertensive medications on measured blood
  pressure. \emph{American journal of hypertension}, \textbf{18}~(7), 935--942.

\bibitem[{Zheng \bbland{} van~der Laan(2012)}]{zheng2012}
Zheng, W. \bbland{} van~der Laan, M.J. (2012) Causal mediation in a survival
  setting with time-dependent mediators. \emph{Technical Report 295, Division
  of Biostatistics, University of California, Berkeley, Calif, USA, 2012.}

\end{thebibliography}

\end{document}